\documentclass[paper]{JHEP3}
\usepackage{booktabs}
\usepackage{psfig}
\usepackage{epsfig,bm,amsmath}
\usepackage{enumerate}
\usepackage{cite}

\preprint{HERWIG/09/02\\
          LU-TP 09-05\\
          MCnet/09/07}
{\title{MC@NLO for the hadronic decay of Higgs bosons in associated production with vector bosons}
\author{Oluseyi Latunde-Dada \\
Dept. Of Theoretical Physics, \\
Solvegatan 14A, S-223 62,\\
Lund, Sweden, \\
E-mail: \email{seyi@hep.phy.cam.ac.uk}}}
\abstract{In this article we describe simulations of the hadronic decay of Higgs bosons produced in
  association with vector bosons at linear and hadronic colliders. We use the Monte Carlo
  at next-to-leading-order ({\tt MC@NLO}) matching prescription with the {\textsf Herwig++}
  event generator to predict various spectra of
  the resulting $b\bar{b}$ pairs and compare our results with leading order and matrix
  element correction predictions.}


\begin{document}
\section{Introduction}
The Higgs boson is an elusive particle which couples to particles according to their mass
and so is weakly coupled to quarks and leptons. The dominant production mechanisms of
Higgs bosons at hadron colliders is from gluon-gluon fusion and vector boson fusion. In
this paper, we consider another mechanism which may be more relevant to an experimental
search. This is the associated production of Higgs bosons with vector bosons. This
production mechanism, $p\bar{p} \rightarrow WH/ZH +X
$, is the most promising discovery channel for a light Standard Model (SM) Higgs boson
at the
Tevatron. This is because the Higgs, which decays predominantly into $b \bar{b}$ pairs, can be tagged by the
associated vector boson.  

Such processes can be simulated in parton shower generators which resum soft and collinear
leading logarithmic, as well as an important subset of next-to-leading logarithmic
contributions to all orders. These simulations can further be improved by matching the parton shower to
higher order matrix elements. One way in which this is done in the generic parton shower
generators is through the use of the matrix element correction \cite{Seymour:1994df} which generates
harder emissions in regions outside the reach of the parton shower at a rate given by the
matrix element. A more rigorous matching procedure is the {\tt MC@NLO} method
\cite{Frixione:2002ik,Frixione:2003ei,Frixione:2005vw,Frixione:2008yi,Frixione:2008ym} which has been implemented for a multitude of processes in the {\tt HERWIG} event
generator \cite{Corcella:2000bw} and for some processes \cite{LatundeDada:2007jg} in its successor the {\textsf Herwig++} event
generator \cite{Bahr:2008pv,Bahr:2008tf}. More recently, another matching method was proposed, called the POsitive Weighted Hardest Emission Generator
({\tt POWHEG}) \cite{Nason:2004rx,Frixione:2007vw}, which achieves the same aim as {\tt MC@NLO}, with the creation of
positive weighted events and is furthermore independent of the shower generator used. The {\tt POWHEG} method
has been applied to $Z$ pair hadroproduction \cite{Nason:2006hfa}, heavy flavour production \cite{Frixione:2007nw},
$e^+e^-$ annihilation to hadrons \cite{LatundeDada:2006gx}, Drell-Yan vector boson production \cite{Alioli:2008gx,Hamilton:2008pd} and
top pair production at the ILC \cite{LatundeDada:2008bv}.

In this paper, we aim to simulate the NLO hadronic decay of the light Higgs boson produced in
association with a vector boson using the {\tt MC@NLO} method. The parton shower generator
we will be employing is {\textsf  Herwig++}. In Section \ref{crosssections}, we first discuss the
cross-sections for associated production of the Higgs boson with a vector boson and its subsequent hadronic
decay rate. We then discuss the application of the {\tt MC@NLO} method to the decay in Section \ref{MCNLO}. In Section \ref{results}, we show
some comparative distributions obtained from the parton shower and in Section
\ref{conc} we summarize our conclusions. Finally, it should be noted that in this paper, we do not apply the {\tt MC@NLO} method
to the initial state emissions.

\section{Cross-sections and decay rates}
\label{crosssections}
\subsection{Associated Higgs production with a $W$ boson from $q\bar{q}$ annihilation}
\label{WHCS} 
\begin{figure}
\begin{center}
\psfig{figure=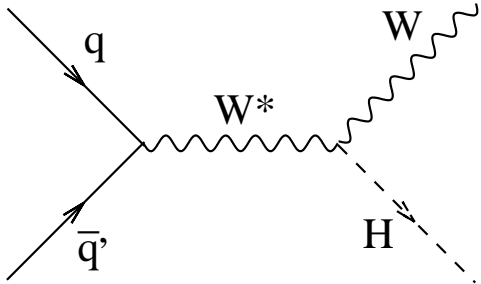,%
width=2.5in,height=1.5in,angle=0}
\end{center}
\caption{Associated $WH$ boson production.}
\label{fig:WH}
\end{figure}

The process $q(p_q) + \bar{q}'(p_{\bar{q}}') \rightarrow W^* \rightarrow W(p_W) + H(p_H)$ is illustrated in Figure \ref{fig:WH}.
If we define the center of mass energy squared of the partonic system by
\begin{eqnarray}
s &=& (p_q +p_{\bar{q}'})^2  \,,
\end{eqnarray}
we have for the differential cross-section \cite{Ciccolini:2003jy},
\begin{eqnarray}
\label{diff}
\frac{d \sigma_p}{d \cos \theta^*} &=& \frac{G_F^2M_W^2 V_{q\bar{q}'}^2}{\cos^2 \theta_W}\frac{\beta_W
  \gamma_W M_W^3}{48 \pi s^{3/2}} \left(\frac{s+M_W^2-M_H^2}{s-M_W^2}\right)^2 [2(1-\beta_W^2) + \beta_W^2 \sin^2
\theta^*] \,,
\end{eqnarray}
where $\theta^*$ is the angle between the $W$ boson and the quark in the partonic CMF,
$G_F$ is the Fermi coupling constant, $\theta_W$ is the Weinberg mixing angle, $M_W$ and $M_H$ are
respectively the $W$ and Higgs boson masses and $V_{q\bar{q}'}$ is the relevant CKM matrix
element. $\beta_W$ is the speed of the $W$ boson in the partonic CMF and is given by
\begin{equation}
\beta_W = \frac{\sqrt{[s-(M_W+M_H)^2][s-(M_W-M_H)^2]}}{s-M_H^2+M_W^2} \;.
\end{equation}
The relativistic boost factor $\gamma_W$ is $(1-\beta_W^2)^{-1/2}$.
Equation \ref{diff} when integrated gives for the total partonic cross-section,
\begin{eqnarray}
\sigma_p &=& \frac{G_F^2M_W^2 V_{q\bar{q}'}^2}{\cos^2 \theta_W}\frac{\beta_W
  \gamma_W M_W^3}{12 \pi s^{3/2}} \left(\frac{s+M_W^2-M_H^2}{s-M_W^2}\right)^2
\left[1-\frac{2 \beta_W^2}{3} \right] \;.
\end{eqnarray}
Convolving this with parton distribution functions (PDFs), we obtain the hadronic
cross-section as
\begin{eqnarray}
\label{sigh}
\sigma_h = \int_{\tau}^{1} dx_1 \int_{\tau/x_1}^{1}
dx_2 [f_q(x_1, M_W^2) f_{\bar{q}'}(x_2, M_W^2)+ x_1 \leftrightarrow x_2] \sigma_p \,,
\end{eqnarray}
where $x_1, x_2$ are the momentum fractions of the incoming partons and taking $S$ as the hadronic beam-beam center-of-mass energy, we have $s = x_1x_2S$.
\subsection{Associated Higgs production with a $Z$ boson from $e^+e^-$ annihilation}
The differential cross-section for the process $e^+e^- \rightarrow Z^* \rightarrow ZH$
\begin{figure}
\begin{center}
\psfig{figure=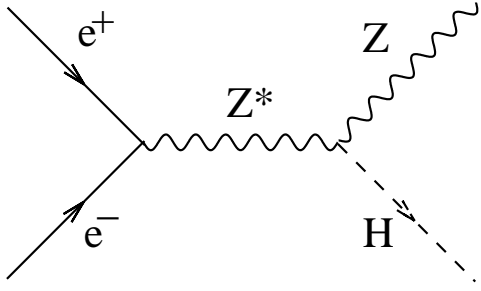,%
width=2.5in,height=1.5in,angle=0}
\end{center}
\caption{Associated $ZH$ production.}
\label{fig:ZH}
\end{figure}
is given by \cite{Mahlon:1998jd}
\begin{eqnarray}
\label{diff2}
\frac{d \sigma}{d \cos \theta^*} &=& \frac{G_F^2M_W^2}{\cos^2 \theta_W} \frac{\beta_Z
  \gamma_Z M_Z^3}{32 \pi s^{3/2}} \left(\frac{s+M_Z^2-M_H^2}{s-M_Z^2}\right)^2 \times
\nonumber \\
&&\left[1-4 \sin^2 \theta_W + 8 \sin^4 \theta_W \right][2(1-\beta_Z^2) + \beta_Z^2 \sin^2
\theta^*] \,,
\end{eqnarray}
where $\theta^*$ is the angle between the electron
and the $Z$ boson in the CMF, $s$ is the
center-of-mass energy and $\gamma_Z$ and $\beta_Z$ are obtained from the analogous
expressions for $W$ production in section \ref{WHCS} by substituting $M_Z$ for $M_W$.  
Equation \ref{diff2} integrated over $\cos \theta^*$ gives
\begin{eqnarray}
\sigma(e^+e^- \rightarrow ZH) &=& \frac{G_F^2M_W^2}{\cos^2 \theta_W} \frac{\beta_Z
  \gamma_Z M_Z^3}{8 \pi s^{3/2}} \left(\frac{s+M_Z^2-M_H^2}{s-M_Z^2}\right)^2 \times
\nonumber \\
&&\left[1-4 \sin^2 \theta_W + 8 \sin^4 \theta_W \right]\left[1 - \frac{2}{3}
  \beta_Z^2\right] \;.
\end{eqnarray}
\subsection{Higgs boson decay to $b\bar{b}$ pairs}
\subsubsection{Lowest Order decay rate}
\begin{figure}
\begin{center}
\psfig{figure=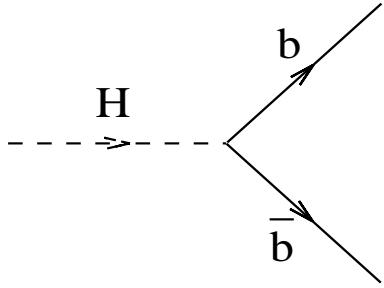,%
width=2in,height=1.5in,angle=0}
\end{center}
\caption{Lowest order Higgs hadronic decay rate.}
\label{fig:Hbb}
\end{figure}
The lowest-order decay rate for this process, illustrated in Figure \ref{fig:Hbb}, is given by
\begin{equation}
\Gamma_B(H \rightarrow b \bar{b}) = \frac{3 G_F m_b^2}{4 \sqrt{2} \pi} M_H \beta_0^3 \,,
\end{equation}
where $m_b$ is the mass of the bottom quark and $\beta_0 = \sqrt{1-\frac{4m_b^2}{M_H^2}}$.
\subsubsection{Virtual radiative corrections}
\begin{figure}
\begin{center}
\psfig{figure=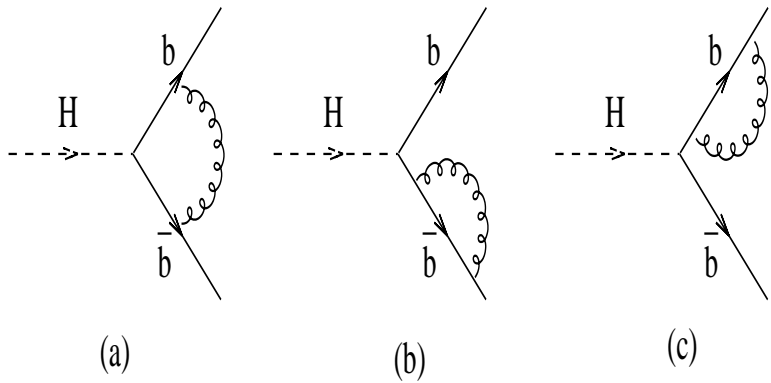,%
width=4in,height=2in,angle=0}
\end{center}
\caption{Virtual corrections (a): vertex correction (b),(c): self-energy corrections.}
\label{fig:virt}
\end{figure}
If one uses the on-shell renormalization scheme, the self-energy diagrams in Figures
\ref{fig:virt}(b) and \ref{fig:virt}(c) are
cancelled by counter-term diagrams leaving us with the vertex correction in Figure
\ref{fig:virt}(a) and its
counter-term. When evaluated in the massive gluon regularization scheme, the final result is
\cite{Drees:1990dq},
\begin{eqnarray}
\label{sigV}
\Gamma_{V} &=& \frac{\alpha_S C_F}{\pi}\Gamma_B \left[-\left(\frac{1+\beta_0^2}{2 \beta_0} \ln
    \frac{1+\beta_0}{1-\beta_0} \right) \ln \frac{m_b^2}{\mu^2}
\right. \nonumber \\
  &+& \left. \frac{1+\beta_0^2}{\beta_0} \left\{{\rm Li}_2\left(\frac{1-\beta_0}{1+\beta_0}\right) +
  \ln \frac{1+\beta_0}{2\beta_0}\ln\frac{1+\beta_0}{1-\beta_0}-\frac{1}{4} \ln^2
  \frac{1+\beta_0}{1-\beta_0} +\frac{\pi^2}{3} \right\} \right. \nonumber \\
&+& \left. \frac{1-\beta_0^2}{\beta_0}\ln\frac{1+\beta_0}{1-\beta_0}-1\right] \,,
\end{eqnarray}
where we have introduced the gluon mass $\mu$ to regulate the infrared singularity in
Figure \ref{fig:virt}(a).
\subsubsection{Real emission corrections}
\begin{figure}
\begin{center}
\psfig{figure=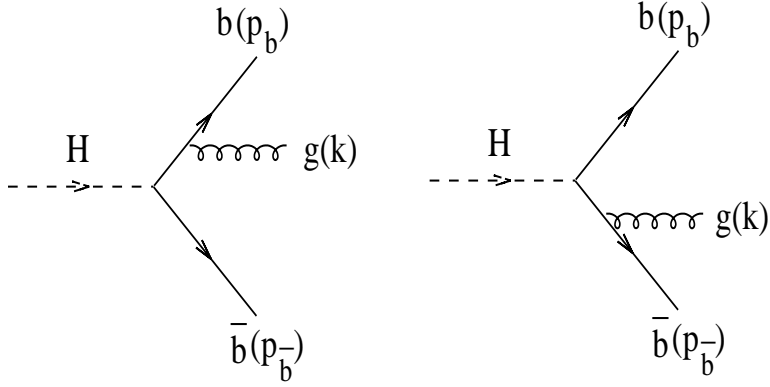,%
width=4in,height=2in,angle=0}
\end{center}
\caption{Real gluon emission.}
\label{fig:real}
\end{figure}
Using the following relations for the energy fractions of the $b$ and $\bar{b}$ quarks in terms of the parton momenta in 
the Higgs rest frame,
\begin{eqnarray}
x_b &=& 1 - \frac{2p_{\bar{b}} \cdot k}{M_H^2} \,, \nonumber \\
x_{\bar{b}} &=& 1 -\frac{2p_b \cdot k}{M_H^2} \,,
\end{eqnarray}
we have for the real emission decay rate, the following expression:
\begin{eqnarray}
\label{dsigR}
\frac{d\Gamma_R}{dx_b dx_{\bar{b}}} &=& \Gamma_B \frac{\alpha_S C_F}{2\pi}\mathcal{M} \nonumber \\
&=&\Gamma_B\frac{\alpha_S C_F}{2\pi
  \beta_0^3} \left[\frac{(1-x_b)^2+(1-x_{\bar{b}})^2 + 2(1-2\rho)(x_b+x_{\bar{b}} -1 - 4
    \rho)}{(1-x_b)(1-x_{\bar{b}})} \right. \nonumber \\
&+& \left. 4\rho\left \{\frac{1}{1-x_b}+\frac{1}{1-x_{\bar{b}}}
  \right\}-2\rho(1-4\rho)\left \{\frac{1}{(1-x_b)^2}+\frac{1}{(1-x_{\bar{b}})^2} \right\}
  +2 \right] \,,   
\end{eqnarray}
where $\rho = \frac{m_b^2}{M_H^2}$.
This can be integrated in the massive gluon scheme to get,
\begin{eqnarray}
\label{sigR}
\Gamma_R &=& \Gamma_B\frac{\alpha_S C_F}{2\pi}\left[\left(\frac{1+\beta_0^2}{2
      \beta_0}\ln\frac{1+\beta_0}{1-\beta_0} \right) \ln \frac{m_b^2}{\mu^2} \right.\nonumber \\
&+&\left. \frac{1+\beta_0^2}{2\beta_0}\left\{3{\rm
      Li}_2\left(\frac{1-\beta_0}{1+\beta_0}\right)+2{\rm
      Li}_2\left(-\frac{1-\beta_0}{1+\beta_0}\right) -
    \ln\frac{2}{1+\beta_0}\ln\frac{1+\beta_0}{1-\beta_0} \right. \right. \nonumber \\
&+& \left. \left. \frac{1}{4}\ln^2\frac{1+\beta_0}{1-\beta_0} + \ln
    \frac{1+\beta_0}{2\beta_0}\ln\frac{1+\beta_0}{1-\beta_0} -\frac{\pi^2}{3} \right \}
  -3\ln\frac{4}{1-\beta_0^2} -4\ln \beta_0 \right. \nonumber \\
&+&\left. \frac{1}{16 \beta_0^3}\left \{3 + 2\beta_0^2 +3
    \beta_0^4\right\}\ln\frac{1+\beta_0}{1-\beta_0}+\frac{1}{8\beta_0^2}\{-3+29
  \beta_0^2\} \right] \;. 
\end{eqnarray}
Summing this with $\Gamma_B$ and $\Gamma_V$ in equation \ref{sigV}, the dependence of the gluon mass
$\mu$ disappears to give,
\begin{equation}
\Gamma_{\rm NLO} = \Gamma_B\left[1 + \frac{\alpha_SC_F}{\pi}\Pi\right] \,,
\end{equation}
where
\begin{eqnarray}
\Pi &=& \frac{1}{\beta_0}\left[(1+\beta_0^2) \left \{4{\rm
      Li}_2\left(\frac{1-\beta_0}{1+\beta_0}\right)+ 2{\rm
      Li}_2\left(-\frac{1-\beta_0}{1+\beta_0}\right) -
    3\ln\frac{2}{1+\beta_0}\ln\frac{1+\beta_0}{1-\beta_0}\right. \right.\nonumber \\ 
&-& \left. \left.2\ln
    \beta_0\ln\frac{1+\beta_0}{1-\beta_0} \right\} -3\beta_0\ln\frac{4}{1-\beta_0^2} - 4
  \beta_0\ln\beta_0 \right] \nonumber \\
&+& \frac{1}{16 \beta_0^3}[3+34 \beta_0^2 -13
  \beta_0^4]\ln\frac{1+\beta_0}{1-\beta_0}+\frac{3}{8 \beta_0^2}[-1+7 \beta_0^2] \;.
\end{eqnarray}
Also from equations \ref{sigV}, \ref{dsigR} and \ref{sigR}, we note that $\Gamma_V$ can be written in
terms of the real emission matrix element squared $\mathcal{M}$ as 
\begin{eqnarray}
\label{dsigV}
\Gamma_V &=& \Gamma_B\frac{\alpha_S C_F}{2\pi}\left[\left\{-\int dx_b dx_{\bar{b}}  \hspace{0.1in} \mathcal{M}\right\} +
  2\Pi_V \right] \,,
\end{eqnarray}
where
\begin{eqnarray}
\Pi_V &=& \frac{1-\beta_0^2}{\beta_0}\ln\frac{1+\beta_0}{1-\beta_0} +
\frac{1+\beta_0^2}{\beta_0}\left\{4{\rm
      Li}_2\left(\frac{1-\beta_0}{1+\beta_0}\right)+ 2{\rm
      Li}_2\left(-\frac{1-\beta_0}{1+\beta_0}\right) \right. \nonumber \\
&+&\left. 2\ln\frac{1+\beta_0}{2\beta_0}\ln\frac{1+\beta_0}{1-\beta_0} - \ln\frac{2}{1+\beta_0}
\ln\frac{1+\beta_0}{1-\beta_0} \right\}  - 3\ln\frac{4}{1-\beta_0^2}-4 \ln\beta_0
\nonumber \\
&+&\frac{1}{16 \beta_0^3}\{3+18 \beta_0^2 +3
  \beta_0^4\}\ln\frac{1+\beta_0}{1-\beta_0}+\frac{3}{8 \beta_0^2}\{-1+7 \beta_0^2\} \;.
\end{eqnarray}
\section{{\tt MC@NLO} method}
\label{MCNLO}
\begin{figure}
\begin{center}
\psfig{figure=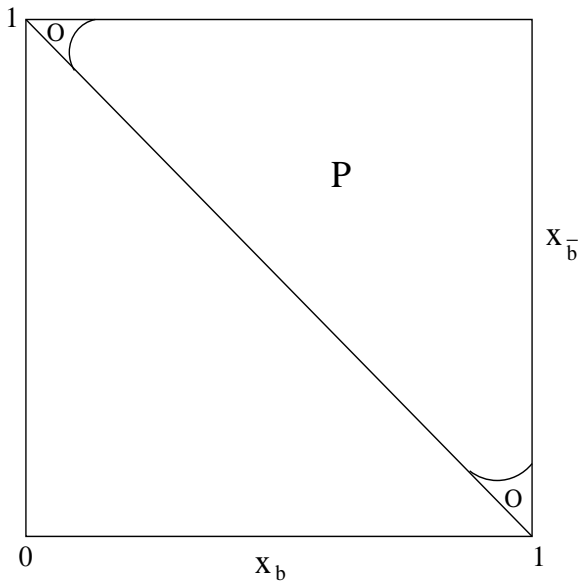,%
width=3in, height=3in, angle=0}
\end{center}
\caption{Phase space for gluon emission.}
\label{fig:ps1}
\end{figure}

The phase space for gluon emission is given in terms of the Dalitz plot variables $x_b,
x_{\bar{b}}$ by 
\begin{equation}
\lambda(x_b^2-\rho, x_{\bar{b}}^2-\rho, (2-x_b-x_{\bar{b}})^2) \leq 0 \,,
\end{equation}
where the function $\lambda$ is defined by,
\begin{equation}
\label{lambda}
\lambda(x,y,z) = x^2 +y^2+z^2 -2xy-2yz-2xz \;.
\end{equation}
This is equivalent to the condition
\begin{equation}
(1-x_b)(1-x_{\bar{b}})(x_b+x_{\bar{b}}-1) > \rho(2-x_b-x_{\bar{b}})^2 \,,
\end{equation}
Figure \ref{fig:ps1} shows the corresponding phase space region
where we have labeled as ${\bf P}$ the emission region and ${\bf O}$ the region outside
the ${\bf P}$ but in the
half-triangle $(1-x_b)(1-x_{\bar{b}})(x_b+x_{\bar{b}}-1) > 0$. 
 
Now using equation \ref{dsigR} and \ref{dsigV}, we can re-write $\Gamma_{\rm NLO}$ in integral form as,
\begin{equation}
\Gamma_{\rm NLO} = \sigma_B \int_{\bf P} dx_b dx_{\bar{b}}  \left[2 - \frac{\alpha_S
    C_F}{2\pi}\{\mathcal{M} - 4\Pi_V\} + \frac{\alpha_S
    C_F}{2\pi}\mathcal{M} \right] + \Gamma_B\int_{\bf O} dx_b dx_{\bar{b}}\left[2 + \frac{\alpha_S
    C_F}{2\pi}4\Pi_V\right] \;.
\end{equation}
Now, if we define a functional $\mathcal{F}_i$ which represents hadronic final states
generated by a parton shower starting from a configuration $i$, we can write down an
overall generating functional for hadrons from Higgs boson decay as
\begin{eqnarray}
\label{func}
\mathcal{F} &=& \Gamma_B \int_{\bf P} dx_b dx_{\bar{b}}  \left[\mathcal{F}_{b\bar{b}} \left\{2 - \frac{\alpha_S
    C_F}{2\pi}(\mathcal{M} - 4\Pi_V)\right \} + \mathcal{F}_{b\bar{b}g}\frac{\alpha_S
    C_F}{2\pi}\mathcal{M} \right] \nonumber \\
&+& \Gamma_B\int_{\bf O} dx_b dx_{\bar{b}}
\mathcal{F}_{b\bar{b}}\left[2 + \frac{\alpha_S C_F}{2\pi}4\Pi_V\right] \,,
\end{eqnarray}
where $\mathcal{F}_{b\bar{b}}$ is the functional representing shower final states
resulting from the process $H \rightarrow b\bar{b}$ and $\mathcal{F}_{b\bar{b}g}$
represents final states from $H \rightarrow b\bar{b}g$. 

There are two problems with this functional as it is written above. The first is the
highly inefficient sampling that will be required to generate starting $b \bar{b}$ and $b \bar{b} g$
configurations according to the $\mathcal{M}$ since it is divergent in the soft and
collinear regions of phase space.  
The second problem arises because when interfaced with the parton shower, leading order configurations starting with $b\bar{b}$
would radiate quasi-collinear gluons with a distribution given by the parton shower
approximation to $\mathcal{M}$ which we shall call $\mathcal{M}_C$. These are already included in the starting $b \bar{b} g$ configurations. Likewise, some of the
$b\bar{b}g$ configurations would include $b\bar{b}$-like configurations if the gluon is
quasi-collinear to either the quark or anti-quark. This problem is often referred to as {\it
  double-counting}.

Before we discuss how to solve these problems, let us investigate
$\frac{\alpha_SC_F}{2 \pi}\mathcal{M}_C$. In the parton shower generator {\textsf Herwig++}, this is the
massive quasi-collinear splitting function for the emission of a gluon from a quark \cite{Catani:1996vz}: 
\begin{equation}
\label{split}
\frac{\alpha_SC_F}{2 \pi}\mathcal{M}_C d\tilde{q}^2 dz =  \frac{\alpha_SC_F}{2 \pi}\frac{d\tilde{q}^2}{\tilde{q}^2}\frac{dz}{1-z}\left[1+z^2
  -\frac{2m_b^2}{z \tilde{q}^2}\right] \;.
\end{equation} 
Here $z$ and $\tilde{q}$ are {\textsf Herwig++} evolution variables and are respectively the light-cone
momentum fraction of the quark after the emission of the gluon and an
angular variable related to the relative transverse momentum of the quark after the
emission \cite{Gieseke:2003rz}.
The splitting function in equation \ref{split} for the $b$ quark can be re-written in terms of the Dalitz plot
variables as
\begin{equation}
\frac{\alpha_SC_F}{2 \pi}\mathcal{M}_C dx_b dx_{\bar{b}} = \frac{\alpha_SC_F}{2 \pi}
\frac{dx_b
  dx_{\bar{b}}}{(1-x_{\bar{b}})\sqrt{x_{\bar{b}}^2-4\rho}}\left[\frac{1+z^2}{1-z}-\frac{2\rho}{1-x_{\bar{b}}}\right] \,,
\end{equation}
 where if $r$ is defined as
\begin{equation}
r = \frac{1}{2} \left(1+\frac{\rho}{1+\rho-x_{\bar{b}}} \right) \,,
\end{equation}
then $z$ is given by
\begin{equation}
z = r + \frac{x_b -(2-x_{\bar{b}})r}{\sqrt{x_{\bar{b}}^2-4\rho}} \;.
\end{equation}
Note that $x_b, x_{\bar{b}}$ are given in terms of $z$, $r$ and $\tilde{q}$ by
\begin{eqnarray}
x_b &=& 1-z(1-z)\frac{\tilde{q}^2}{M_H^2} \,, \nonumber \\
x_{\bar{b}} &=& (2-x_b)r+(z-r)\sqrt{x_b^2 - 4\rho} \;.
\end{eqnarray}
For emission from $\bar{b}$ anti-quark interchange $x_b$ and $x_{\bar{b}}$ in the equations
above.

\begin{figure}
\begin{center}
\psfig{figure=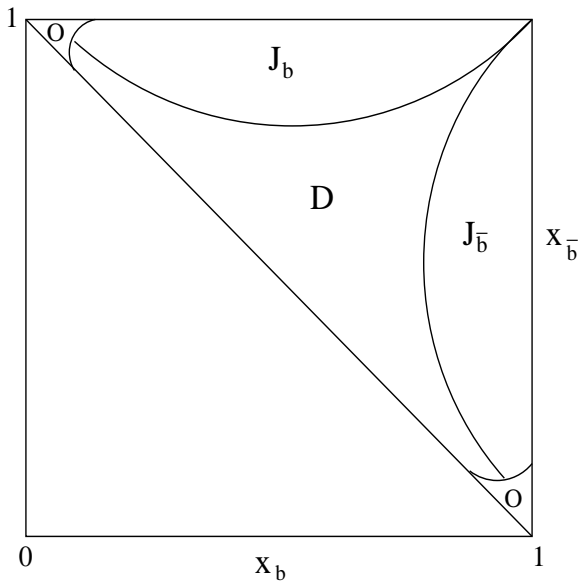,%
width=3in, height=3in, angle=0}
\end{center}
\caption{Phase space showing hard ${\bf D}$ and soft/collinear $ {\bf J_b}, {\bf
    J_{\bar{b}}}$ gluon emission regions.}
\label{fig:ps2}
\end{figure}

In {\textsf Herwig++}, the quasi-collinear region of phase space covered by the parton
shower is defined by imposing the condition,
\begin{equation}
\tilde{q}^2 < \frac{M_H^2}{2}(1+\sqrt{1-4\rho}) \;.
\end{equation}
The corresponding regions are shown in Figure \ref{fig:ps2} labeled ${\bf J_b}, {\bf J_{\bar{b}}}$ whilst the
unpopulated dead region is labeled ${\bf D}$. Note that region ${\bf P}$ in Figure
\ref{fig:ps1} corresponds to the union of regions ${\bf\
 J_b}$, ${\bf J_{\bar{b}}}$ and ${\bf D}$.
Now going back to the functional defined in equation \ref{func}, if in
the region ${\bf J} = {\bf J_{b}} \cup {\bf J_{\bar{b}}}$, we subtract the parton shower approximation
$\mathcal{M}_C$ from the second term in equation \ref{func} and add it to the first term we
get 
\begin{eqnarray}
\label{func2}
\mathcal{F} &=& \Gamma_B \int_{\bf J} dx_b dx_{\bar{b}}  \left[\mathcal{F}_{b\bar{b}}
  \left\{2 - \frac{\alpha_S
    C_F}{2\pi}(\mathcal{M}-\mathcal{M}_C - 4\Pi_V)\right \} + \mathcal{F}_{b\bar{b}g}\frac{\alpha_S
    C_F}{2\pi}\{ \mathcal{M}- \mathcal{M}_C\} \right] \nonumber \\
&+& \Gamma_B \int_{\bf D} dx_b dx_{\bar{b}}  \left[\mathcal{F}_{b\bar{b}}
  \left\{2 - \frac{\alpha_S
    C_F}{2\pi}(\mathcal{M} - 4\Pi_V)\right \} +
\mathcal{F}_{b\bar{b}g}\frac{\alpha_S
    C_F}{2\pi} \mathcal{M} \right] \nonumber \\
&+& \Gamma_B\int_{\bf O} dx_b dx_{\bar{b}}
\mathcal{F}_{b\bar{b}}\left[2 + \frac{\alpha_S C_F}{2\pi}4\Pi_V\right] \;.
\end{eqnarray}
This is the {\tt MC@NLO} method which solves the problem of double-counting in the parton
shower regions ${\bf J}$. It also solves the problem of the sampling inefficiency since $\mathcal{M}
\rightarrow
\mathcal{M}_C$ in the divergent regions and therefore $\mathcal{M}
-\mathcal{M}_C$ tends to $0$ there.
In Appendix \ref{HM}, we describe the algorithm used for the evaluation of the above
integrals and the generation of events. We also discuss how we regularize some residual
divergences by the use of mappings in Appendix \ref{section4}. 

The procedure followed for event generation for associated Higgs production is outlined below.
\begin{enumerate}
\item
For $p\bar{p}$ annihilation use equation \ref{sigh} to distribute the Mandelstam variables $x_1,
x_2$ and the angle $\theta^*$ according to the differential cross-section. From these variables, reconstruct
the $W$ and Higgs boson four-momenta. For $e^+e^-$ annihilation, use equation \ref{diff2} to
distribute the angle $\theta^*$ and reconstruct the $Z$ and Higgs boson four-momenta.
\item
In the rest frame of the Higgs boson, generate the Dalitz plot variables $x_b,
x_{\bar{b}}$ and event weight as described in Appendix \ref{HM}. Reconstruct the four-momenta of the quark,
anti-quark and gluon in this frame.
\item
Boost the four-momenta back to the lab frame.
\end{enumerate}
More details about the {\tt MC@NLO} method can be found in \cite{Frixione:2002ik}.
\section{Results}
\label{results}
\begin{figure}[!ht]
\vspace{0.5cm}
\hspace{0.5cm}
\psfig{figure=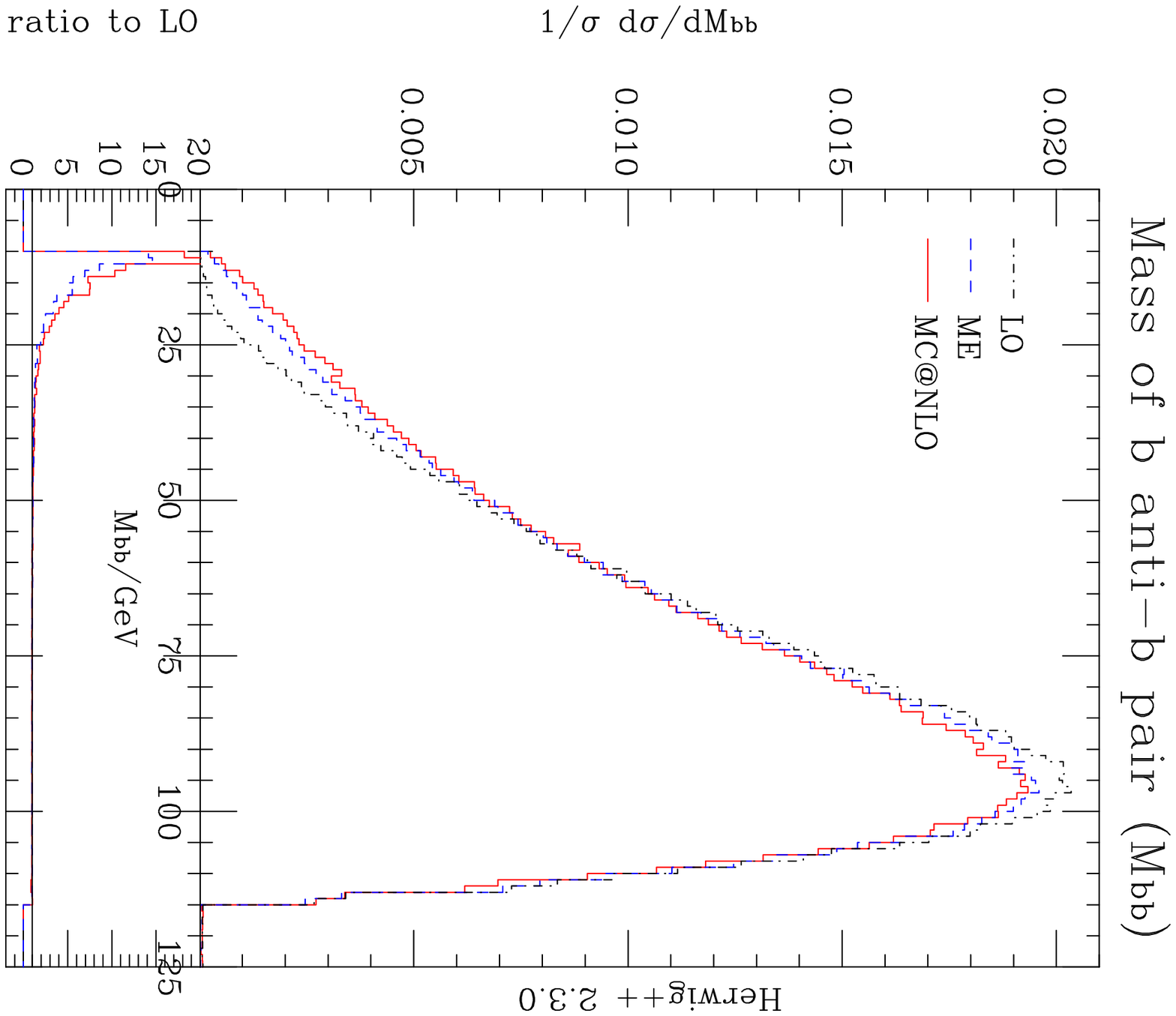,%
width=3in,height=3in,angle=90}
\psfig{figure=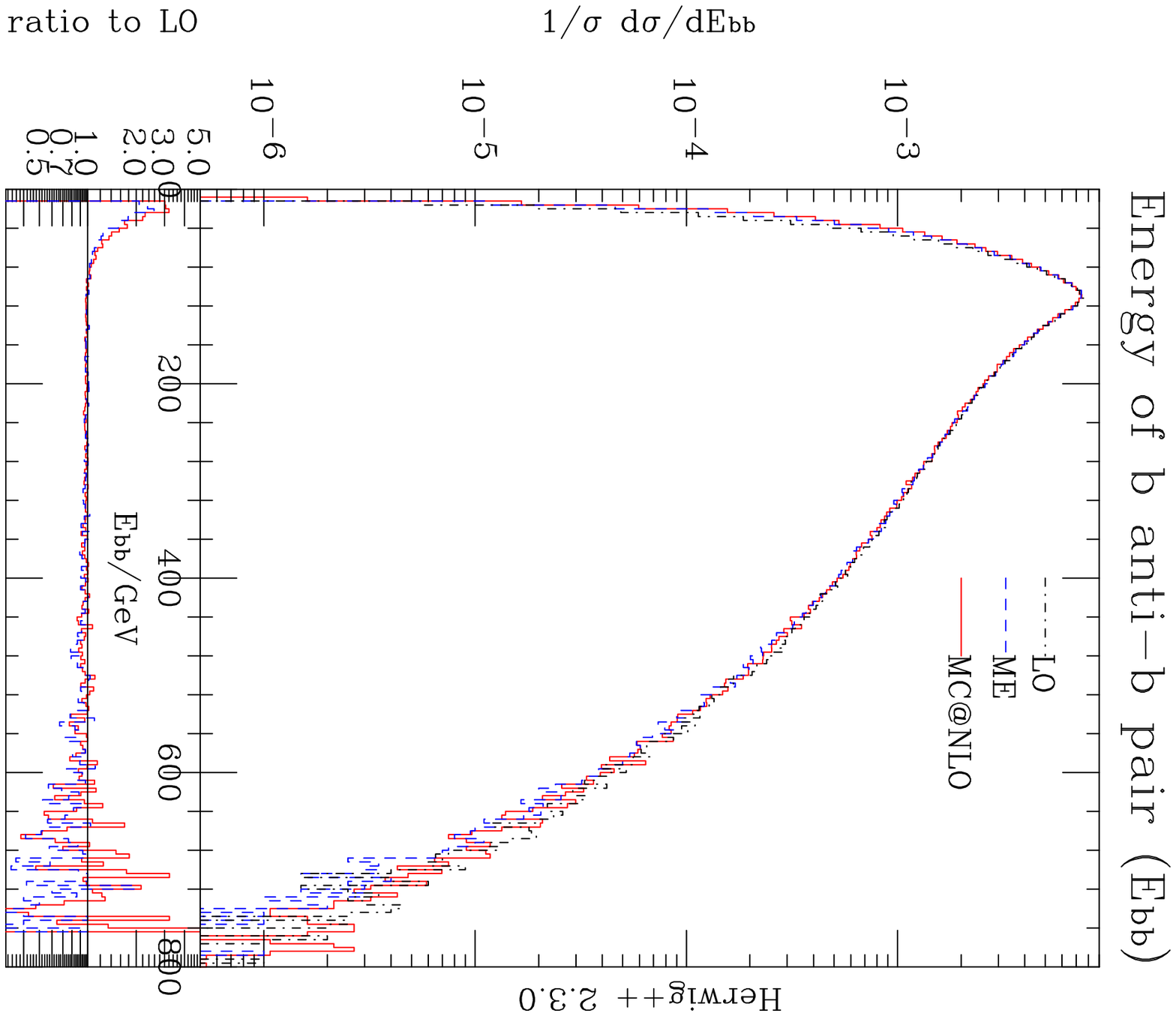,%
width=3in,height=3in,angle=90}
\caption{Mass and energy of the $b\bar{b}$ pair ($q\bar{q}'$ annihilation).}
\label{fig:masspp}
\end{figure}

\begin{figure}[!ht]
\vspace{0.5cm}
\hspace{0.5cm}
\psfig{figure=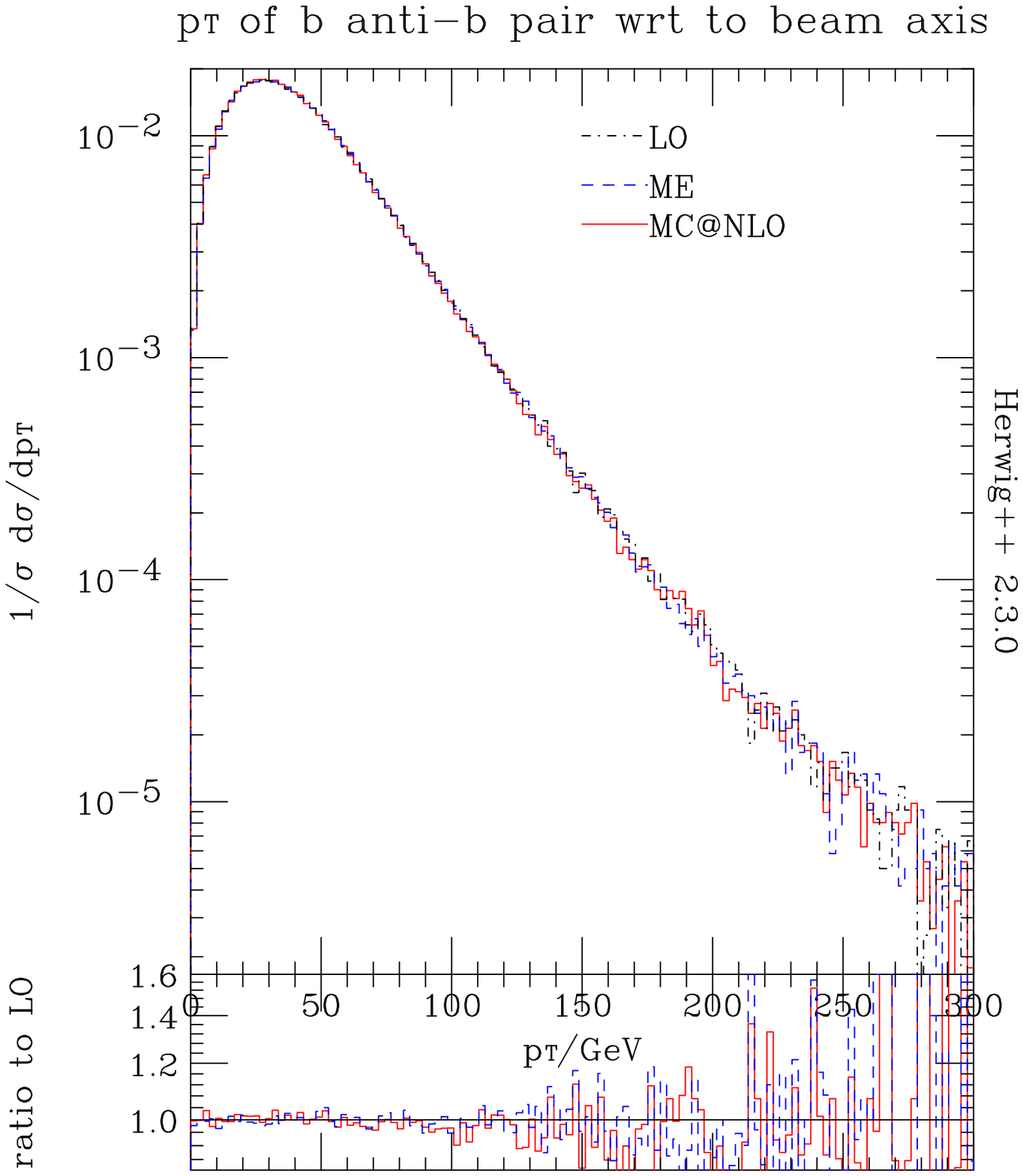,%
width=3in,height=3in,angle=90}
\psfig{figure=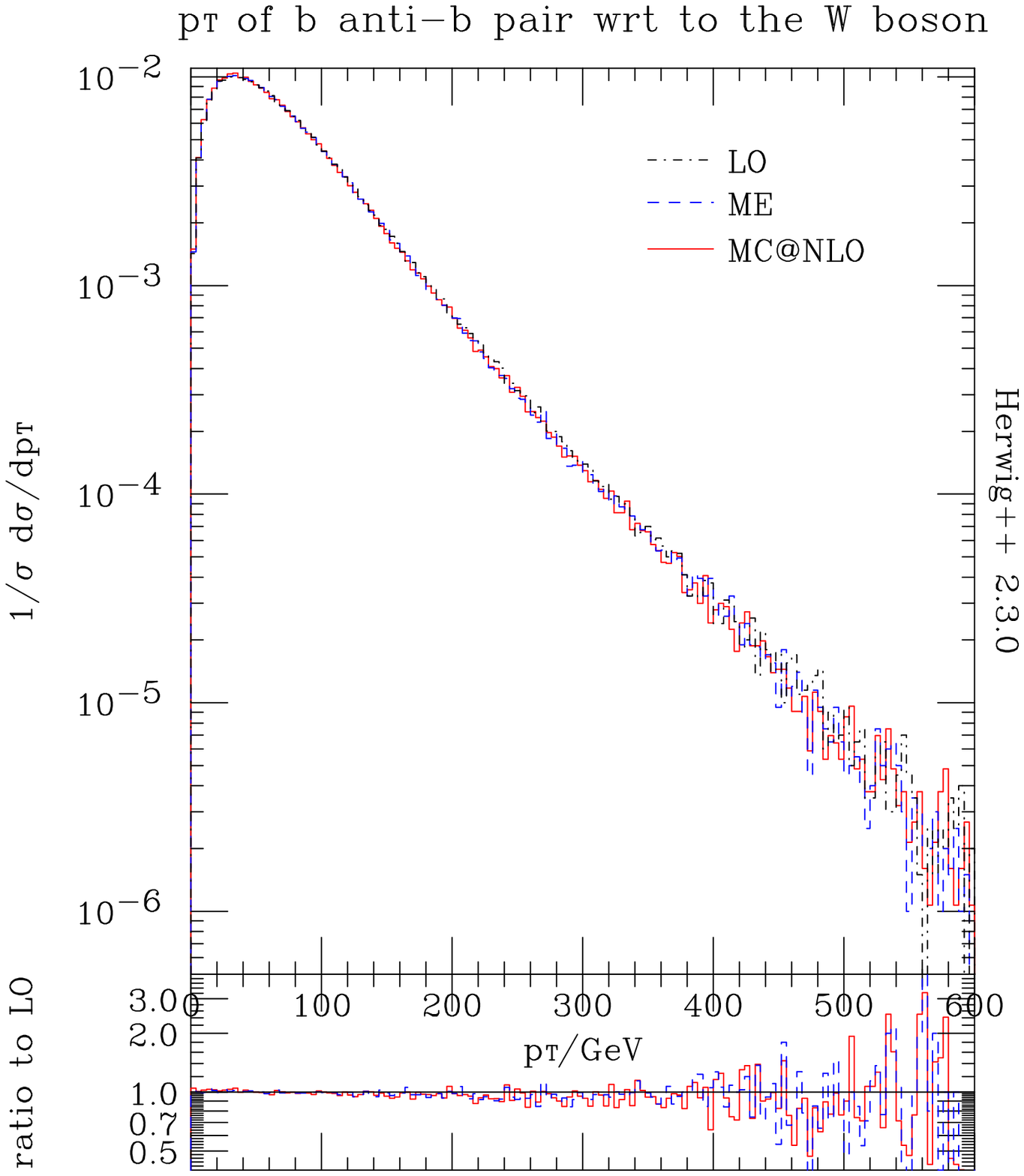,%
width=3in,height=3in,angle=90}
\caption{$p_{\rm T}$ of the $b\bar{b}$ pair w.r.t the beam axis and w.r.t the $W$ boson ($q\bar{q}'$ annihilation).}
\label{fig:ptbpp}
\end{figure}

\begin{figure}[!ht]
\vspace{0.5cm}
\hspace{0.5cm}
\psfig{figure=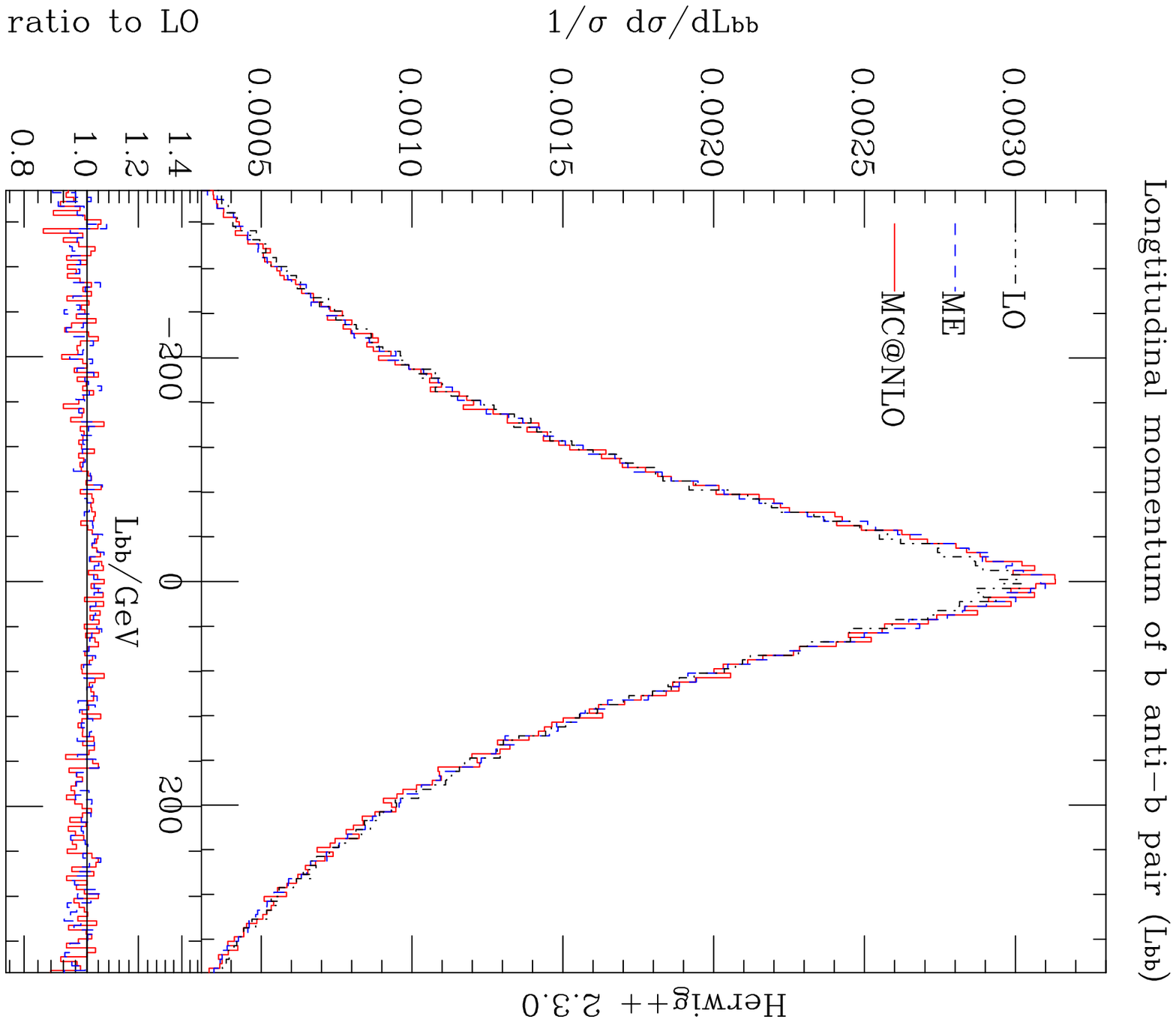,%
width=3in,height=3in,angle=90}
\psfig{figure=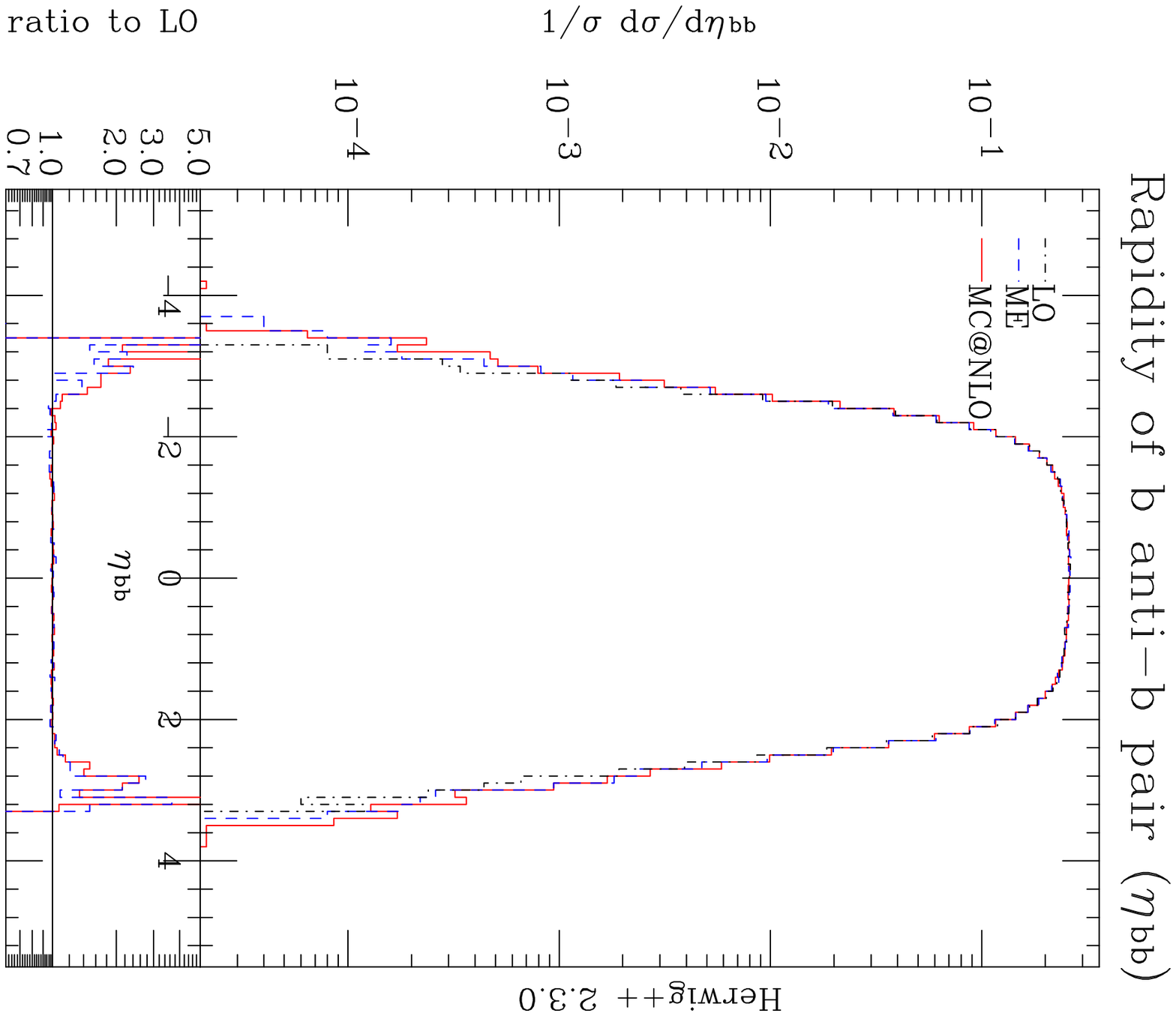,%
width=3in,height=3in,angle=90}
\caption{Longitudinal momentum and rapidity of the $b\bar{b}$ pair ($q\bar{q}'$ annihilation).}
\label{fig:longtpp}
\end{figure}

\begin{figure}[!ht]
\vspace{0.5cm}
\hspace{0.5cm}
\psfig{figure=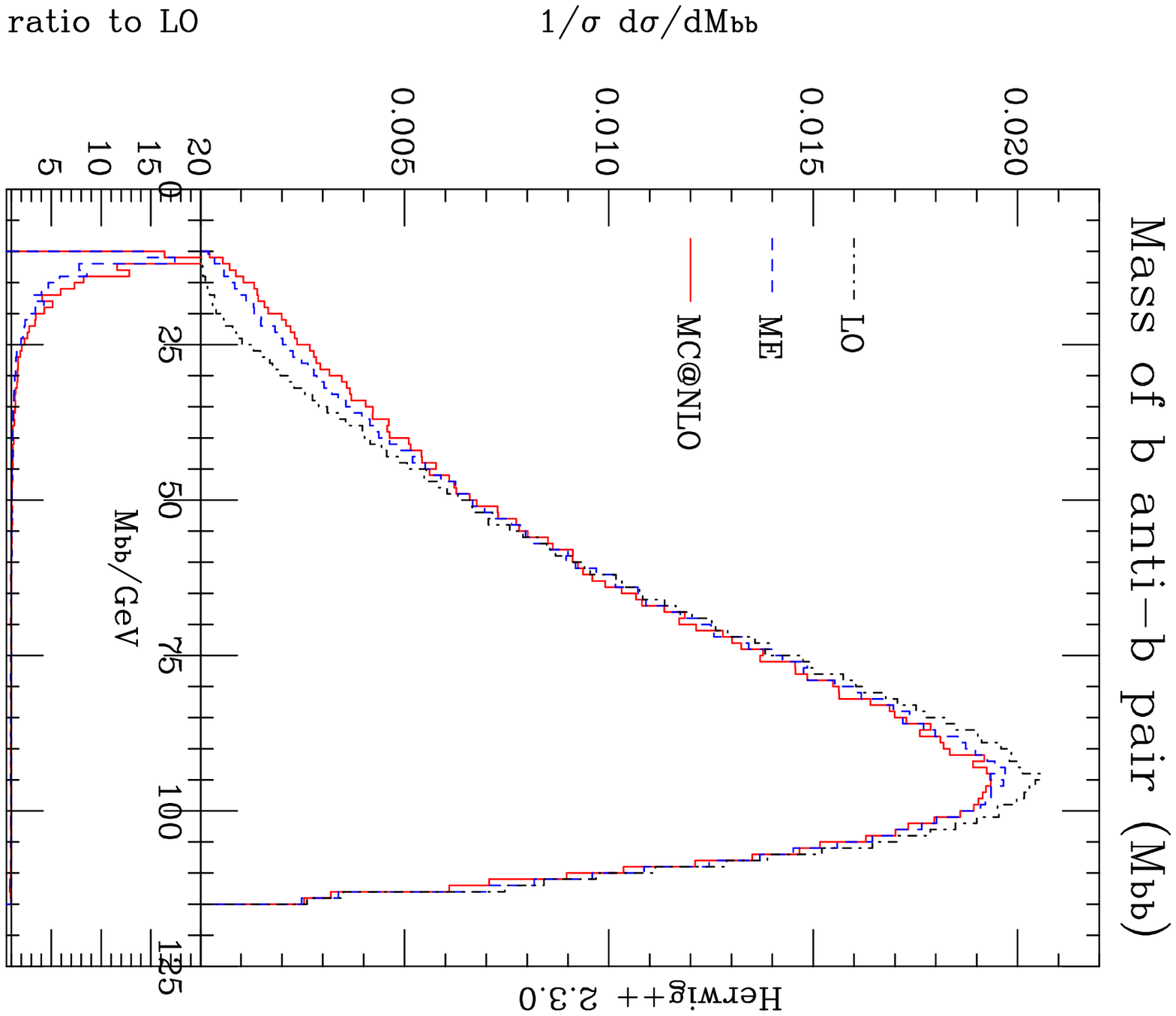,%
width=3in,height=3in,angle=90}
\psfig{figure=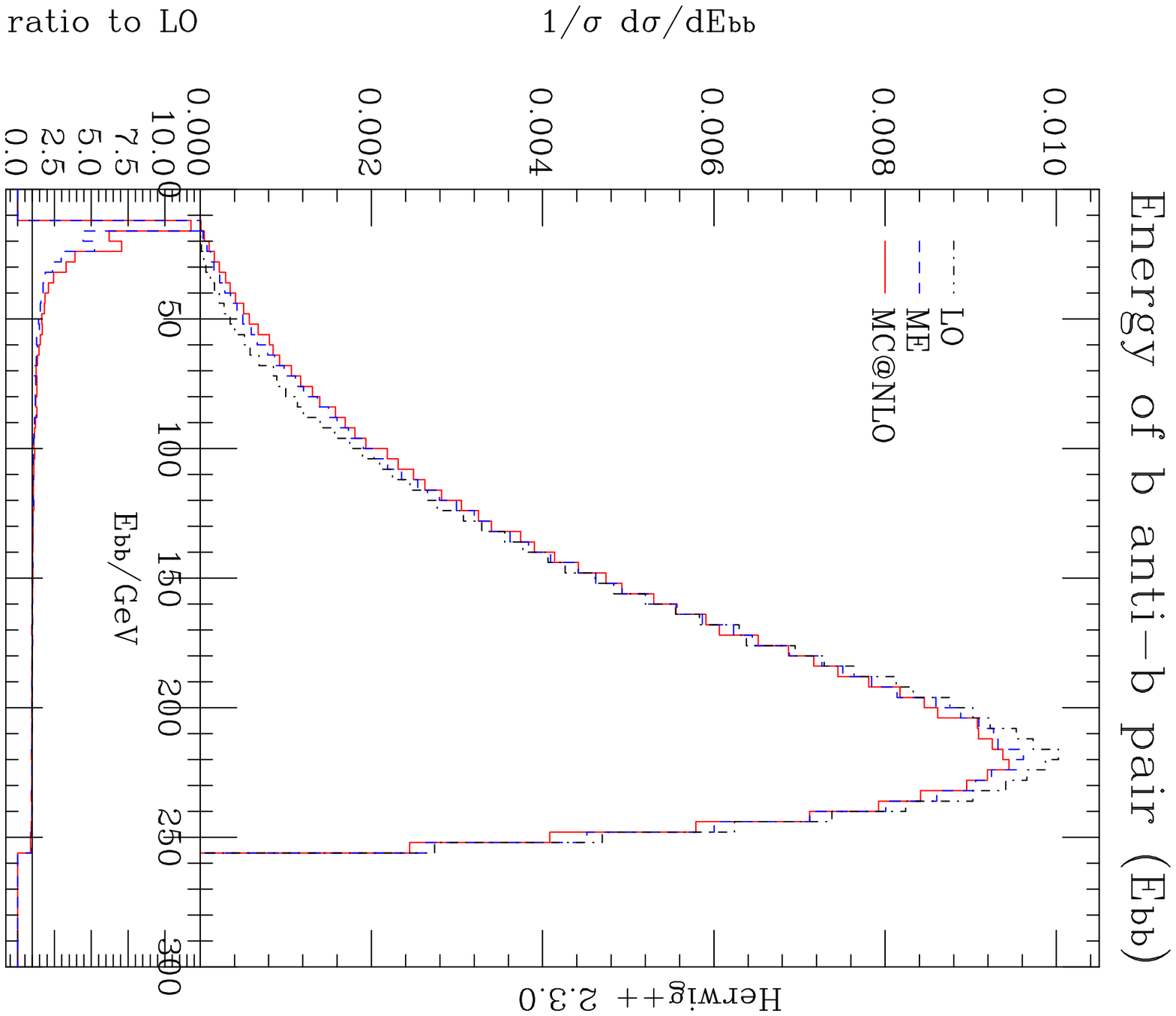,%
width=3in,height=3in,angle=90}
\caption{Mass and energy of the $b\bar{b}$ pair ($e^+e^-$ annihilation).}
\label{fig:massee}
\end{figure}

\begin{figure}[!ht]
\vspace{0.5cm}
\hspace{0.5cm}
\psfig{figure=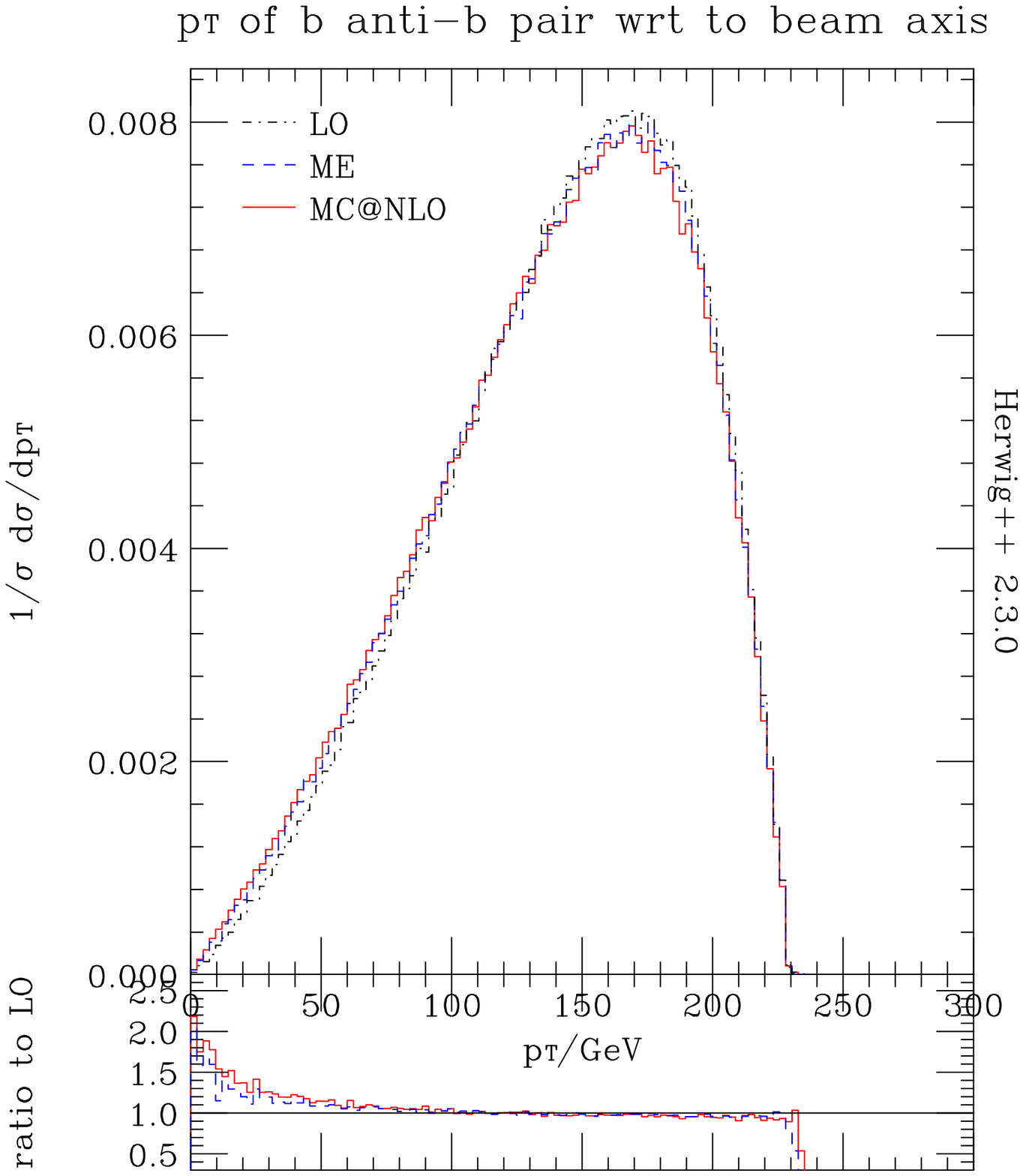,%
width=3in,height=3in,angle=90}
\psfig{figure=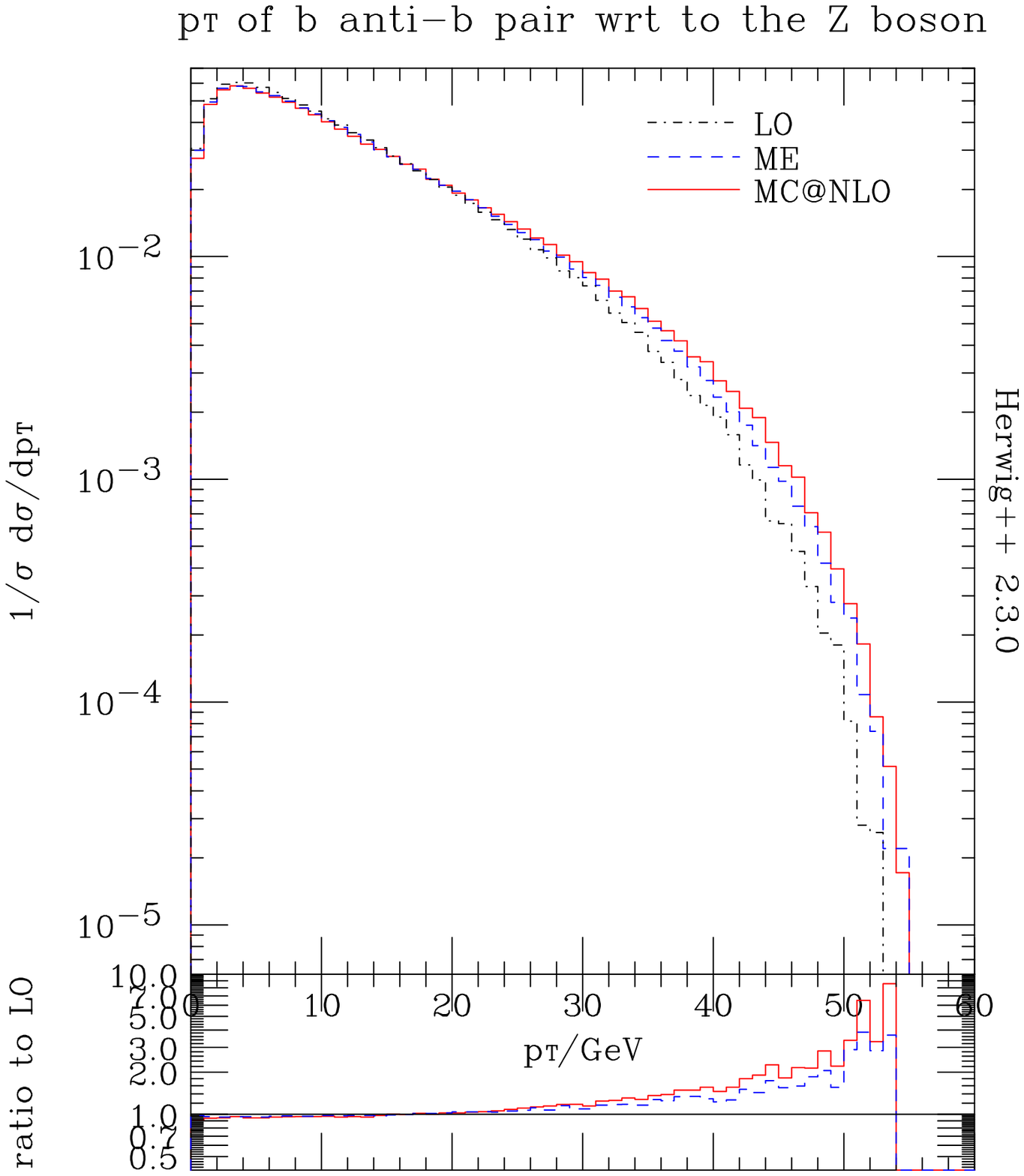,%
width=3in,height=3in,angle=90}
\caption{$p_{\rm T}$ of the $b\bar{b}$ pair w.r.t the beam axis and w.r.t the $Z$ boson ($e^+e^-$ annihilation).}
\label{fig:ptbee}
\end{figure}

\begin{figure}[!ht]
\vspace{0.5cm}
\hspace{0.5cm}
\psfig{figure=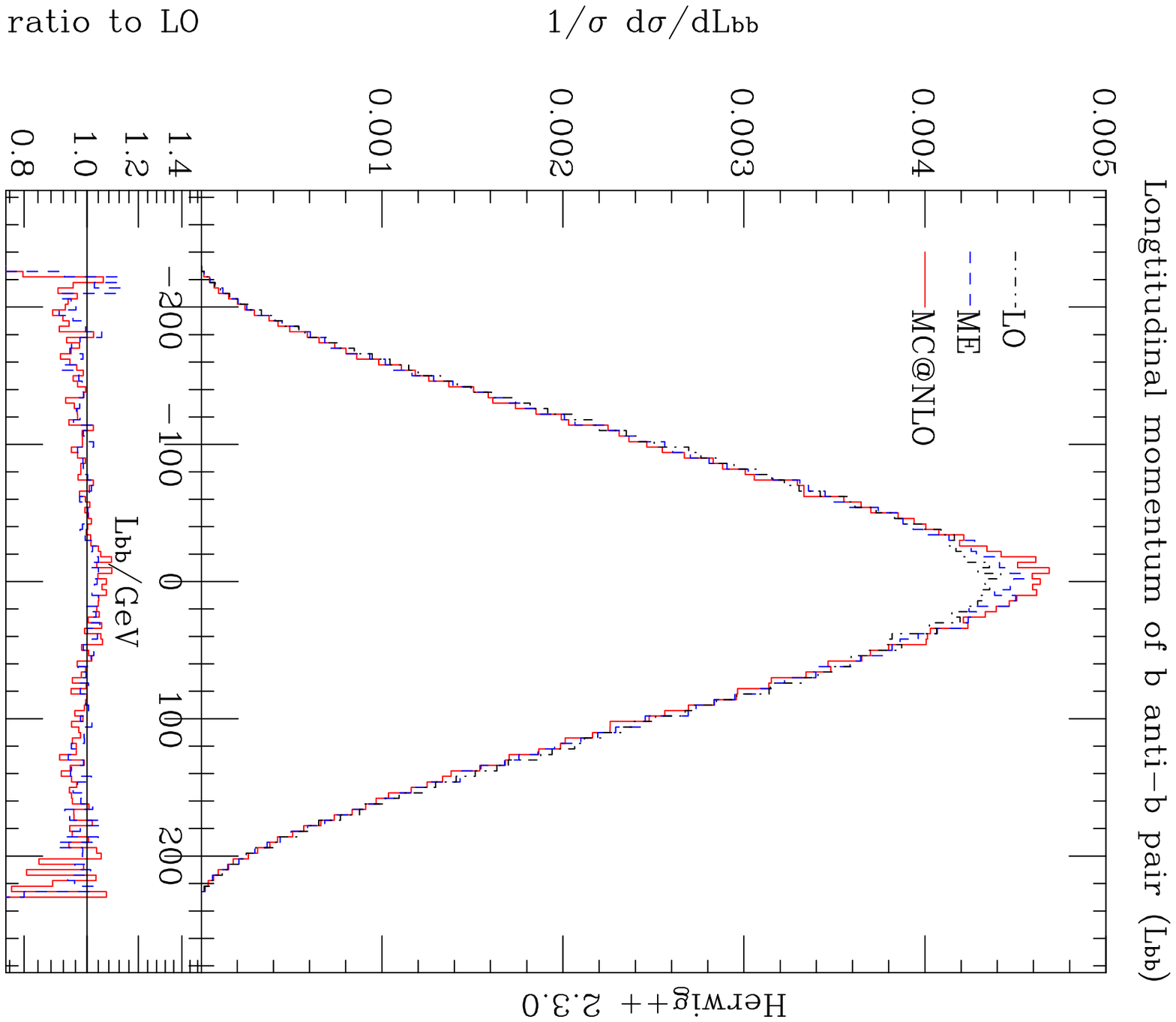,%
width=3in,height=3in,angle=90}
\psfig{figure=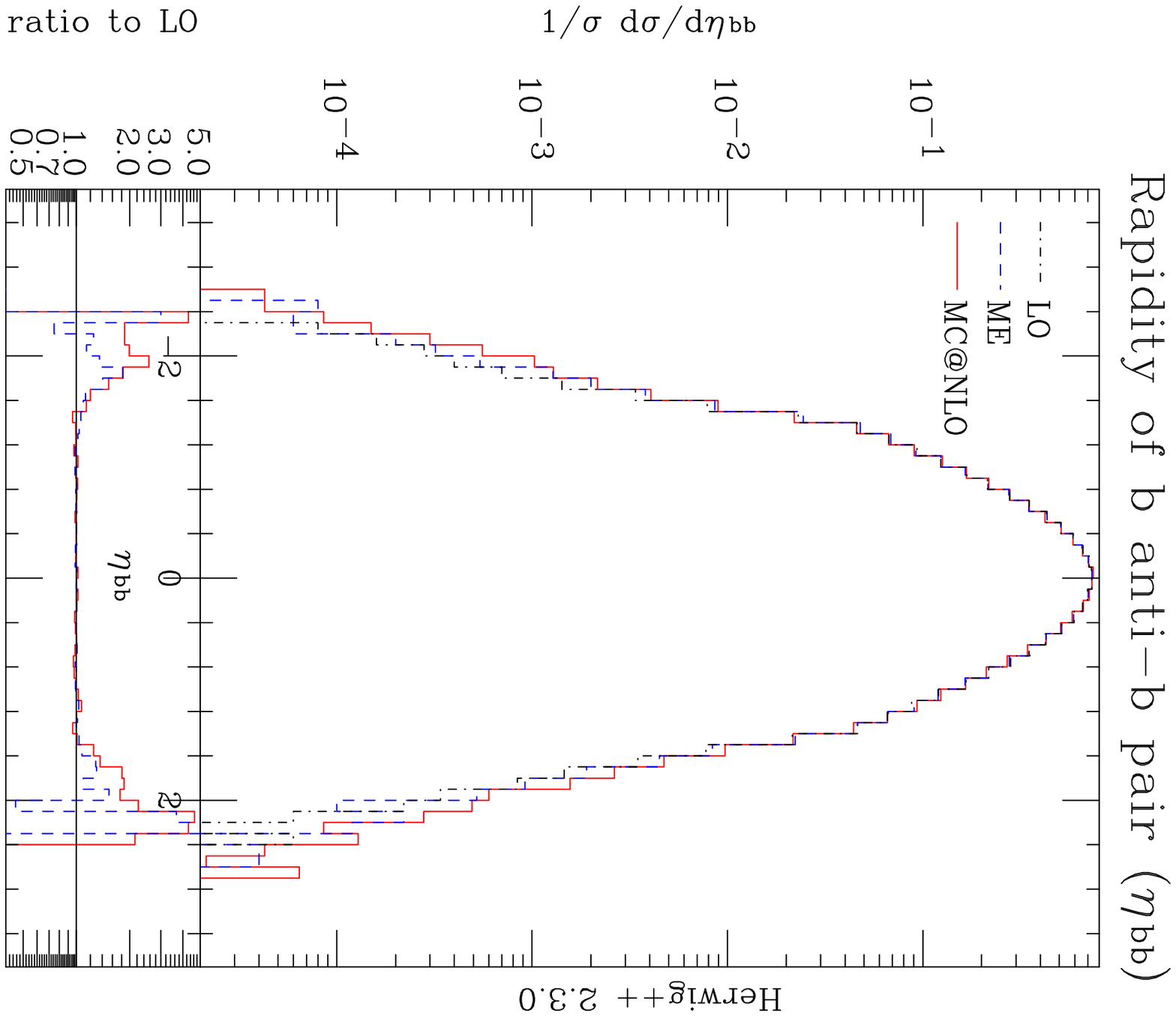,%
width=3in,height=3in,angle=90}
\caption{Longitudinal momentum and rapidity of the $b\bar{b}$ pair ($e^+e^-$ annihilation).}
\label{fig:longtee}
\end{figure}

Following the prescription above, events for associated Higgs boson production of mass
$114$ GeV and decaying into $b\bar{b}$ pairs with $m_b = 5$ GeV,  were
generated and interfaced with {\textsf Herwig++ 2.3.0} \cite{Bahr:2008tf}. The following approximations
were considered:
\begin{enumerate}
\item The {\textsf Herwig++} parton shower interfaced to leading order events (LO),
\item the parton shower interfaced to leading order events which are supplemented by matrix
  element corrected events in the dead region (ME), 
\item the parton shower interfaced to events generated by the {\tt MC@NLO} method (MC@NLO).
\end{enumerate}
The two processes considered were:
\begin{enumerate}
\item associated $WH$ production from $q\bar{q}'$ annihilation at the Tevatron (1.96 TeV),
\item associated $ZH$ production from $e^+e^-$ annihilation at the ILC (0.5 TeV).
\end{enumerate}
The following distributions were considered and shown in Figures \ref{fig:masspp} - \ref{fig:longtpp} for $q\bar{q}'$
annihilation and associated $W$ boson production and Figures \ref{fig:massee} - \ref{fig:longtee} for $e^+e^-$
annihilation and associated $Z$ boson production. In the simulations, only the leptonic
decays of vector bosons were considered.
\begin{enumerate}
\item The mass of the $b \bar{b}$ pair before hadronization,
\item the energy of the $b \bar{b}$ pair,
\item the transverse momentum of the $b\bar{b}$ pair with respect to the beam axis, 
\item the transverse momentum of the $b\bar{b}$ pair with respect to the direction of the
  vector boson,
\item the longitudinal momentum of the $b \bar{b}$ pair with respect to the beam axis,
\item the rapidity of the $b \bar{b}$ pair.
 \end{enumerate}
From the mass and energy reconstruction plots in Figures \ref{fig:masspp} and \ref{fig:massee}, we see that the extra gluon radiation simulated by the {\tt
  MC@NLO} and matrix element correction methods smooth out the respective peaks due to the
production of more low mass and energy $b\bar{b}$ pairs. We also observe as expected that the
matrix element correction method underestimates the amount of hard gluon radiation when
compared to the {\tt MC@NLO} distributions.

For $q\bar{q}'$ annihilation, the effect
of this extra radiation on the transverse momenta plots is diminished by the initial
state radiation from the incoming partons as can be seen in Figure \ref{fig:ptbpp}. The effect is
more clearly seen in the corresponding plots for $e^+e^-$ annihilation in Figure \ref{fig:ptbee}
where there is no initial state radiation. At leading order, the $b\bar{b}$ pairs are
produced predominantly at right angles to the beam axis (equation \ref{diff2}), hence the peak
at high $p_{\rm T}$ in the LO plot for $e^+e^-$ annihilation. The effect of extra gluon
emission is to smooth out the peak as seen in the {\tt MC@NLO} and ME correction plots. Likewise
at leading order, the $b\bar{b}$ pairs are produced back-to-back with the associated $Z$
boson. The effect of gluon radiation is therefore to increase the transverse momentum of
the $b\bar{b}$ with respect to the $Z$ boson. This is what is observed in the {\tt MC@NLO}
and ME correction distributions with the latter underestimating the amount of radiation in the tails in
comparison with the former.

Finally, in Figures \ref{fig:longtpp} and \ref{fig:longtee}, the {\tt MC@NLO} and ME correction method predict
slightly more peaked distributions for the longitudinal
momenta around the
central value of $0$ GeV. This is a result of the loss of energy from the $b\bar{b}$
pairs due to extra gluon radiation. The rapidity
distributions show the {\tt MC@NLO} and ME
correction methods predict the production of more high rapidity $b\bar{b}$ pairs in
comparison to the leading order prediction. This arises as a result of more $b\bar{b}$ pairs being
produced at low $p_{\rm T}$ with respect to the beam axis (Figure \ref{fig:ptbee}) and therefore higher absolute rapidity. It should also be noted that for
both the longitudinal momenta and rapidity predictions, the ME correction
method underestimates the amount of radiation in the tails of the distributions in
comparison to the {\tt MC@NLO} plots.
\section{Conclusions}
\label{conc}
In this work we have realized the {\tt MC@NLO} matching prescription for the decay of 
Higgs bosons produced in association with vector bosons at both the ILC and hadron
colliders. This work was achieved within the framework of the {\textsf Herwig++} Monte Carlo
event generator.

We compared the {\tt MC@NLO} predictions with those obtained via the matrix element
correction method as well as leading order predictions. The effects of the hard
radiation are visible in the reconstruction plots for the mass and energy of the
$b\bar{b}$ pairs resulting from Higgs decay. Also visible is the effect on the
longitudinal momenta and rapidity of the $b\bar{b}$ pairs. 

Less visible is the effect on the $p_{\rm T}$
spectra for $q\bar{q}'$ annihilation due to the dominant effects of initial state
radiation. For $e^+e^-$ annihilation, the {\tt MC@NLO} method predicts a softer spectrum
for the $p_{\rm T}$ with respect to the beam axis and a harder spectrum with respect to the
associated $Z$ boson.

The algorithm used for these processes will be publicly available with the forthcoming
version of {\textsf Herwig++}. Although we have considered the associated production of
Higgs and vector bosons in this
paper, the {\tt MC@NLO} algorithm can be interfaced with other Higgs boson production
mechanisms, since here we only apply the method to the decay process.

\section{Acknowledgements}
We are grateful to the other members of the {\textsf Herwig++} collaboration for
developing the program that underlies the present work and for helpful comments. We are particularly grateful to
Bryan Webber for constructive comments and discussions throughout. This work was supported
by the UK Science and Technology Facilities Council, formerly the Particle Physics and
Astronomy Research Council, and the European Union Marie Curie Research Training Network
MCnet under contract MRTN-CT-2006-035606.
\clearpage
\appendix
\section{Monte Carlo algorithm}
\label{HM}
The integrals in (\ref{func2}) can be evaluated using a variety of Monte Carlo methods. In
this paper, the `Hit or Miss' Monte Carlo method is used. This is the simplest and
oldest form of Monte Carlo integration and essentially involves finding the area of a
region in phase space by integrating over a larger region, a binary function which is $1$ in
the region and $0$ elsewhere.  The sampling method used for the points $x_b,x_{\bar{b}}$ is
the importance sampling method whereby more samples are taken from regions where the
integrand is large and less from regions where it is small. This ensures that the sampled
points have the same distribution as the integrand.

The following algorithm summarizes how the starting $b\bar{b}$ and $b \bar{b} g$
configurations were generated according to equation \ref{func2}. In the discussion that
follows, we work in the rest frame of a Higgs boson of mass $M_H = 114$ GeV and
$\alpha_S = \alpha_S(M_H)=0.114$.
\begin{enumerate}
\item
Randomly sample points $x_b,x_{\bar{b}}$, in each of regions
${\bf J_{b}},{\bf J_{\bar{b}}}$, ${\bf D}$ and ${\bf O}$ of the phase space
and using the `Hit Or Miss' Monte Carlo method, evaluate the 5 integrals,
$I_{J}^{(2)},I_{J}^{(3)},I_{D}^{(2)}$, $I_{D}^{(3)}$ and $I_{O}^{(2)}$ as well as their absolute sum,
$I$.
\begin{eqnarray}
\label{eq13}
I_{J}^{(2)}&=&\int_{\bf J}dx_b
dx_{\bar{b}}\left[2-\frac{\alpha_S}{2\pi}C_F\left\{\mathcal{M}-\mathcal{M}_{C}-\Pi_V \right\}\right] \,,\nonumber\\
I_{J}^{(3)}&=&\int_{\bf J}dx_b dx_{\bar{b}}\frac{\alpha_S}{2\pi}C_F[\mathcal{M}-\mathcal{M}_{C}]\,, \nonumber\\
I_{D}^{(2)}&=&\int_{\bf D}dx_b
dx_{\bar{b}}\left[2-\frac{\alpha_S}{2\pi}C_F\left\{\mathcal{M}-\Pi_V\right\}\right]\,,\nonumber\\
I_{D}^{(3)}&=&\int_{\bf D}dx_b dx_{\bar{b}}\frac{\alpha_S}{2\pi}C_F\mathcal{M} \,,\nonumber\\
I_{O}^{(2)}&=&\int_{\bf O}dx_b
dx_{\bar{b}}\left[2+\frac{\alpha_S}{2\pi}C_F\Pi_V \right] \,,\nonumber\\
I&=&\mid{I_{J}^{(2)}}\mid+\mid{I_{J}^{(3)}}\mid+\mid{I_{D}^{(2)}}\mid+\mid{I_{D}^{(3)}}\mid
+ \mid{I_{O}^{(2)}}\mid \;.
\end{eqnarray}
Note also the maximum values of the integrands in $I_{J}^{(3)}$ and $I_{D}^{(3)}$.
\item
The eventual proportion of $b\bar{b}$ Monte Carlo events will be determined by the ratio $\frac
{\mid{I_{J}^{(2)}}\mid+\mid{I_{D}^{(2)}}\mid + \mid{I_{O}^{(2)}}\mid}{I}$. Likewise, the
proportion of $b \bar{b}g$
events in the soft regions
${\bf J_{b}},{\bf J_{\bar{b}}}$ and the hard region ${\bf D}$ are determined by the ratios
$\frac{\mid{I_{J}^{(3)}}\mid}{I}$ and $\frac{\mid{I_{D}^{(3)}}\mid}{I}$ respectively. The algorithm
below is then used to importance-sample the $b\bar{b}g$ events so that the corresponding
($x_{b},x_{\bar{b}}$) values of the Monte Carlo events have the same distribution as the
integrands in $I_{J}^{(3)}$ and$I_{D}^{(3)}$:
\begin{enumerate}
\item
For event generation in region ${\bf L}$ (where ${\bf L}$ is one of ${\bf D},{\bf J_b}$ or ${\bf J_{\bar{b}}}$), randomly select a point
$x_b,x_{\bar{b}}$ in that region.\\
\item
Evaluate the absolute value of the  integrand in $I_{L}^{(3)}$ for this point,
$\mid{w(x_b,x_{\bar{b}})}\mid$.\\
 Is $\mid {w(x_b,x_{\bar{b}})}\mid$ $> R$ $\mid{w_{\rm max}}\mid$ ? ($R$ is a random
 number between 0
 and 1 and $\mid {w_{\rm max}}\mid$ is the maximum value of
 $\mid{w(x_b,x_{\bar{b}})}\mid$ determined in Step 1).\\
\item
If NO, return to (a). If YES, accept the event and set
$w^{\rm unw}$= sgn $w(x_b,x_{\bar{b}})$ i.e. $w^{\rm unw} = 1$ if $w(x_b,x_{\bar{b}})$ is
positive and $-1$ if
negative. (In regions ${\bf J_{b}}$ and ${\bf J_{\bar{b}}}$, $\mathcal{M} <
\mathcal{M}_{C}$ and hence the integrands and the
integral, $I_{J}^{(3)}$ in these regions are negative). This process is called {\tt
  unweighting}.\\
\item
Repeat the process until the correct proportion of $b\bar{b}$ and $b\bar{b}g$ events have been
generated.\\
\item
Using the importance-sampled points, obtain an estimate for the integral,
$I_L^{(2,3)}=\frac{\sum{w^{\rm unw}}}{N}\times I$, where $N$ is the total number of Monte
Carlo
events generated. We typically use $N=10^6$.\\
\end{enumerate}
\end{enumerate}
This is the {\tt MC@NLO} method. In this way, for a  total of $N$ events, the correct proportion of $b\bar{b}$
and $b\bar{b}g$
events with $\pm$ unit weight is generated  with the same distribution as the integrands
in
(\ref{eq13}). All of these integrals are finite, but the integrands are divergent at
isolated points within the integration regions. Before the sampling could
be carried out, the divergences  in the integrands (which cause problems in the sampling
process) had to be taken care of. This is the described in section \ref{section4}.

\section{Divergences and mappings}
\label{section4}
\subsection{Divergences in dead region}
\begin{figure}[h!]
\begin{center}
\[
\psfig{figure=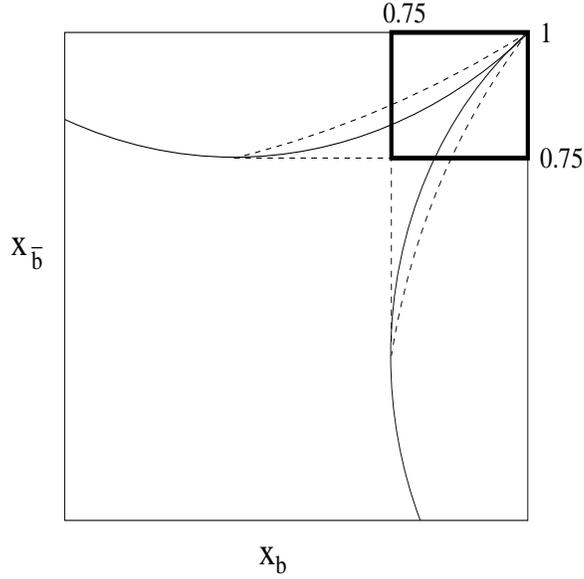,%
width=3in,height=3in,angle=0}
\]
\caption{The mapped region (solid) and the region mapped onto (dashed)}
\label{mapC1}
\end{center}
\end{figure}

In region ${\bf D}$, the hard matrix element squared $\mathcal{M}$ given in equation \ref{dsigR},
diverges as $(x_b,x_{\bar{b}}) \rightarrow (1,1)$. To avoid this
divergence, one can map the divergent region into another region in such a way that the
divergence is regularized. This is ensured by the fact that the region of integration
vanishes as the singularity is approached. There is a double pole in $\mathcal{M}$ at $(x_b,x_{\bar{b}})=(1,1)$. To avoid this pole, the region
$x_b,x_{\bar{b}}>\frac{3}{4}$ is mapped into a region which includes ${\bf D}$ but whose width
vanishes quadratically as $x_b,x_{\bar{b}}\rightarrow 1$ \cite{Gieseke:2003rz, LatundeDada:2007jg}. The mapping is:
\begin{eqnarray}
x_b^{'}&=&1-\left[\frac{1}{4}-(1-x_b)\right]=\frac{7}{4}-x_b \,,\nonumber \\
x_{\bar{b}}^{'}&=&1-2(1-x_b^{'})\left[\frac{3}{4}-(1-x_{\bar{b}})\right]=\frac{5}{8}+\frac{1}{2}x_b+\frac{3}{2}x_{\bar{b}}-2x_bx_{\bar{b}}
\label{eqC1}
\end{eqnarray}
when $x_b>x_{\bar{b}}>\frac{3}{4}$. This mapping also introduces an extra weight factor of
$2(1-x_b^{'})$ in the integrand. Interchange $x_b$ and $x_{\bar{b}}$ in both the mapping
and weight factor when $x_{\bar{b}}>x_b>\frac{3}{4}$. Figure \ref{mapC1} shows the region
mapped (solid) and the region mapped onto (dashed).
\subsection{Divergences in jet regions \boldmath{$J_{b}$} and \boldmath{$J_{\bar{b}}$}}
\label{AB}
\begin{figure}[h!]
\begin{center}
\[
\psfig{figure=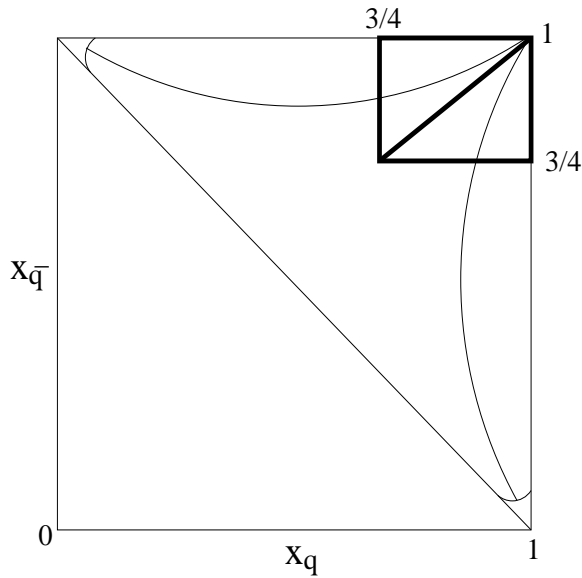,%
width=3in,height=3in,angle=0}
\]
\caption{Mapped regions}
\label{mapAB}
\end{center}
\end{figure}

In both regions ${\bf J_{b}}$ and ${\bf J_{\bar{b}}}$, there is a simple pole in the term $(\mathcal{M}-\mathcal{M}_{C})$
at
$(x_b,x_{\bar{b}})=(1,1)$. In the region $x_b,x_{\bar{b}}>\frac{3}{4}$, a new set of
random points are generated which have a weight factor to cancel the divergence. The
mapping used in region ${\bf J_{b}}$ where $x_{\bar{b}}>x_b$ is \cite{LatundeDada:2007jg}:
\begin{eqnarray}
x_b^{'}&=&1-0.25r_1 \,,\nonumber \\
x_{\bar{b}}^{'}&=&1-(1-x_b^{'})r_2
\label{eqB}
\end{eqnarray}
where $r_1$ and $r_2$ are random numbers in the range $[0,1]$. The weight factor for this
mapping is $2r_1$. For region ${\bf J_{\bar{b}}}$, where $x_b>x_{\bar{b}}$, interchange $x_b$
and
$x_{\bar{b}}$ in the mapping. The mapped regions are shown with solid boundaries in Figure
\ref{mapAB}.
\clearpage
\bibliography{thesis}
\bibliographystyle{utphys}
\end{document}